\RequirePackage[cmex10]{amsmath}
\documentclass{llncs}

\usepackage[pdftex]{graphicx}
\usepackage{algorithmic}
\usepackage{array}
\usepackage{url}
\usepackage{bm}
\usepackage{booktabs}
\usepackage{colortbl}
\usepackage{xcolor}
\usepackage{makeidx}
\usepackage{multirow}

\usepackage[utf8]{inputenc}

\begin{document}

\pagestyle{headings}

\setlength{\belowcaptionskip}{-10pt}

\title{Toward Transparent Heterogeneous Systems}

\author{Baptiste Delporte, Roberto Rigamonti, Alberto Dassatti}

\institute{Reconfigurable and Embedded Digital Systems Institute – REDS HEIG-VD\\School of Business and Engineering Vaud\\HES-SO, University of Applied Sciences Western Switzerland}
\maketitle

\begin{abstract}
  Heterogeneous parallel systems are widely spread nowadays.
  Despite their availability, their usage and adoption are still limited, and even more rarely they are used to full power.
  Indeed, compelling new technologies are constantly developed and keep changing the technological landscape,
  but each of them targets a limited sub-set of supported devices, and nearly all of them require new programming paradigms
  and specific toolsets.
  Software, however, can hardly keep the pace with the growing number of computational capabilities, and developers are less and less
  motivated in learning skills that could quickly become obsolete.

  In this paper we present our effort in the direction of a transparent system optimization
  based on automatic code profiling and Just-In-Time compilation, that resulted in a
  fully-working embedded prototype capable of dynamically detect computing-intensive code blocks
  and automatically dispatch them to different computation units.

  Experimental results show that our system allows gains up to $32\times$ in performance
  --- after an initial warm-up phase --- without requiring any human intervention.
\end{abstract}

\section{Introduction}
\label{sec:Introduction}
\vspace{-1em}
Improvements in computational power marked the last six decades and represented the major factor that allowed mankind to tackle
problems of growing complexity.
However, owing to physical and technological limitations, this process came to an abrupt halt in the past few years~\cite{Lopez15,Keckler11,Asanovic06}.
Industry tried to circumvent the obstacle by switching the paradigm, and both \emph{parallelism} and \emph{specialization} became the keywords
to understand current market trends: the former identifies the tendency of having computing units that are composed of many independent entities
that supposedly increase by a multiplicative factor the throughput; the latter reflects the drift toward architectures (think of, for instance, DSPs)
that are focused on solving particular classes of problems that arise when facing a specific task.
These two non-exclusive phenomena broadened the panorama of technological solutions available to the system developer but, contrary to expectations,
were not capable of sustaining the growth that was observed in the previous years~\cite{Keckler11,Womble99}.
Indeed, while architectures and systems evolve at a fast pace, software does not~\cite{Brooks95}.
Big software projects, which are usually the most computing-intensive ones, demand careful planning over the years, and cannot sustain rapid twitches to
adapt to the latest technological trend.
Developers are even more of a ``static'' resource, in the sense that they require long time periods before becoming proficient in a new paradigm.
Moreover, the most experienced ones, which are the most valuable resource of a software company and those who would be leading the change, are even less
inclined to impose radical turns to a project, as this would significantly affect their mastery and their control of the situation.
Faced with such a dilemma, the only reasonable solution seems to be \emph{automation}.

In this paper we present a solution to this problem capable of detecting segments of code that are \emph{``hot''} from a computational stance and dynamically dispatching
them to different computing units, relieving thus the load of the main processor and increasing the execution speed by exploiting the peculiarities of those
computing units.
In particular, as a case study, we demonstrate our approach by providing a fully-working embedded system, based on the REPTAR platform~\cite{Dassatti14}, that automatically
transfers heavy tasks from the board's Cortex-A8 ARM processor to the C64x+ DSP processor that is incorporated in the same DM3730 chip~\cite{TI3730}.
To achieve this goal, we execute the code we want to optimize in the LLVM~\cite{Lattner04} Just-In-Time (JIT) framework, we then identify functions worth optimizing
by using the Linux's perf\_event~\cite{Weaver13}
tool, and finally we dispatch them to the DSP, aiming at accelerating their execution.
We will hereafter refer to our proposal as \emph{Versatile Performance Enhancer} (VPE).
While the performance we obtain is obviously worse than the one we could achieve by a careful handcrafting of the code, we get this result at no
cost for the developer, who is totally unaware of the environment in which the code will be executed.
Also, as the code changes are performed at run-time, they can adapt to optimize particular input patterns --- think of, for instance, a convolution where most of the
kernel's components are zeros ---, further enhancing the performance.
Finally, the system can dynamically react to changes in the context of execution, for example
resources that become available, are upgraded, or experience an hardware failure.

In the following we will first present the current state of the art, and then accurately describe our approach.
\vspace{-1em}

\section{State of The Art}
\label{sec:vpe_sota}
\vspace{-1em}
Parallel and heterogeneous architectures are increasingly widespread. While a lot of research addresses
Symmetrical Multi-Processors (SMP) systems, heterogeneous architectures are yet not well integrated
in the development flow, and integrators usually have to come up with ad-hoc, non-portable solutions.
Slowly the scenario is changing and mainstream solutions start to be proposed. Particularly interesting in this context are some language-based solutions,
such as CUDA~\cite{Danalis10}, OpenCL~\cite{Stone10,Danalis10}, HSA~\cite{Kyriazis13}, and OpenMP Extensions for Heterogeneous Architectures~\cite{White11}.
All of these solutions are
oriented to some specific hardware settings, they are very similar to a GPU in nature, and are constructed in such
a way that hardware manufacturers can keep a strong hold on their technologies.
While these proposals might look as a perfectly reasonable answer to the heterogeneity problem, they share a major drawback: they all require the programmers to learn
a new programming paradigm and a new toolset to work with them.
Moreover, these solutions are similar but mainly incompatible in nature, and this fact worsens even further the situation.
Supporting more than one approach is expensive for companies, and the reality is that the developer has
seldom the choice about the methodology to adopt once the hardware platform is selected --- most likely by the team in charge of the hardware part, who is probably unaware, or partially unaware, of the
implications that choice has on the programming side.

A partial solution to this problem, called SoSOC, is presented in~\cite{Nasrallah13}: to avoid setting the knowledge of the DSP architecture and the related toolset
(compiler, library, \ldots) as an entry barrier to using the DM3730 chip for actual development,
the authors wrote a library that presented a friendly interface to the programmer and allowed the dispatching of functions to a set of targets based on either the developer's wishes
or some statistics computed during early runs.
While interesting and with encouraging results, in our view the approach has a major drawback:
not only the user of SoSOC has to learn (yet) another library, but also someone has to provide handcrafted code
for any specialized unit of interest.
This is a considerable waste of time and resources, and limits the applicability of the system to the restricted subset of architectures directly supported by the development team.
Furthermore, the developer might not be aware of the real bottlenecks of the system for a particular input set, so he might candidate for remote execution
non-relevant functions, wasting precious resources.

Other academic proposals exist.
A notable one is StarPU from INRIA Bordeaux~\cite{Augonnet11}.
StarPU provides an API and a pragma-based environment that, coupled with a run-time scheduler
for heterogeneous hardware, composes a complete solution.
While the main focus of the project is CPU/GPU systems, it could be extended to less standard systems.
As for~\cite{Danalis10,Stone10,Kyriazis13,White11,Augonnet11,Nasrallah13}, StarPU
shows the same limitations: a new set of tools and a new language or API to master.

Compared with these alternatives, our solution does not need application developers to be aware of the optimization steps that will be undertaken,
it does not target a specific architecture, and it does not require any additional step from the developer's side.

Another interesting technique, called BAAR and focused on the Intel's Xeon Phi architecture, is presented in~\cite{Damschen15a,Damschen15b}.
This proposal is similar to ours, in that the code to be optimized is run inside LLVM's Just-In-Time framework, and functions deemed to be best executed on the Xeon Phi are offloaded to a
remote server that compiles them with Intel's compiler and executes them.
However, their analysis step lacks the ``Versatility'' that characterizes our approach: functions are statically analyzed using Polly~\cite{Grosser12}, a state-of-the-art polyhedral optimizer
for automatic parallelization, to
investigate their suitability for remote execution, and if this is the case, they are sent to the remote target.
In our proposal, instead, optimizations are triggered according to an advanced performance analyzer, fitting to the current input set under processing and not to
expected-usage scenarios or other compile-time metrics.
This allows us a fine-grained control on the metric to optimize, the strategy to achieve this optimization, and the best target selection for a given task at any moment during the program's life.
\vspace{-1em}

\section{VPE approach}
\label{sec:vpe_approach}
\vspace{-1em}
The analysis in the previous section highlighted the importance of alternatives able to automate the code acceleration and dispatching steps.
VPE aims at the run-time optimization of a generic code for a specific heterogeneous platform and input data pair, all in a transparent way.
The idea behind it is that the developer just writes the code as if it had to be executed on a standard CPU.
The VPE framework JIT-compiles this code and executes it, collecting statistics at run-time.
When a user function --- system calls are automatically excluded from the analysis --- behaves according to a specific pattern, for instance, is particularly CPU-intensive, VPE acts to
alter the run-time behaviour trying to optimize the execution.
In the case of a CPU-intensive code, this could be the dispatching on a remote target specialized for the type of operations executed.
After a warm-up delay, which can quickly become negligible for a large family of algorithms adopted in both the scientific and industrial settings, the performances result potentially increased.
If this is not the case --- for instance after an abrupt discontinuity in the input data pattern that makes the computation not suitable for the selected remote target ---, VPE can revise its decisions
and act accordingly.

In structuring VPE, similarly to~\cite{Damschen15a}, we have chosen to cast the problem in the LLVM framework~\cite{Lattner08,Lattner11}.
LLVM recently came as an alternative to the widely known GCC compiler, whose structure was deemed to be too intricate to
allow people to easily start contributing to it.
The biggest culprit seemed to be the non-neat separation between the front-end, the optimization, and the back-end steps~\cite{Lattner08}.
LLVM tried to solve this issue by creating an Intermediate Representation (IR) --- an enriched assembly --- that acts as a common language between the different
steps~\cite{Lattner04}.
The advantage here is that each component of the compilation chain can be unaware of the remaining parts and still be capable of doing its job; for instance,
the ARM back-end does not need to know whether the code it is trying to assemble comes from C++ or FORTRAN code, allowing a back-end designer to focus solely on what
this tool is supposed to do.
As a result, a slew of LLVM-based tools came out in the past few years, with remarkable contributions coming from the academic community too, and this fueled
its diffusion.
Among others, LLVM features a Just-In-Time (JIT) compiler (MCJIT) --- which is the core component of our system --- since many years now, whereas GCC introduced it only at the end of 2014~\cite{GCC5}.
Also, a number of tools allowing in-depth code analysis and optimization, such as~\cite{Venkatesh10,Grosser12,Oh13} to cite a few, can be easily integrated, leaving the door open to future extensions.

We have thus started from MCJIT, integrated an advanced profiling technique, and altered its behaviour by acting directly on the code's IR to allow us dynamically
switch functions at will.
We then took an embedded system that suited our needs, and experimentally verified the improvements introduced by our solution.
\vspace{-1em}

\subsection{Profiling}
\vspace{-.6em}
Detecting which function is the best candidate to be speeded-up is a task that can be accomplished neither
at development time, nor at compile time, as it is usually strongly dependent on inputs.
We therefore had to shape our architecture to include a performance monitoring solution, and after considering different alternatives
(such as OProfile~\cite{Cohen03}), we opted for perf\_event~\cite{Weaver13}.
perf\_event gives access to a large number of hardware performance counters, although at a penalty that can reach up to $20\%$ overhead.
In particular, very interesting measures can be acquired, including cache misses, branch misses, page faults, and many others,
leaving the choice about which figure of merit optimize for, to the system engineer.

In this paper we adopt, as the sole performance metrics for selecting which function off-load, the number of
CPU cycles requested for its execution.
Our only optimization strategy  is blind off-loading --- that is, we off-load the candidate function and we observe if this results in a performance improvement, eventually reverting our choice.
It should be noted, however, that large gains could derive by a careful crafting of this optimization step: as an example, one might think of reorganizing
on-the-fly a data structure after figuring out that it is causing too many cache misses~\cite{Chilimbi99}.

While we do claim that having reliable and accurate statistics is vital to devising a clever optimization strategy, and having such a powerful performance
analyzer integrated in our system is surely a strength of our approach, we will not investigate this topic further in this paper.
\vspace{-1em}

\subsection{Function call techniques}
\vspace{-.6em}
Once an ``interesting'' function is detected, we would like to off-load it to another computational unit (we will refer to this computational unit as \emph{remote target} from now on).
For this to happen, we have to transfer all the function's code, parameters, and shared data to the remote target, then give the control to it, wait for the function to
return, and finally grab the returned values.

\begin{figure}[t]
\centering
\includegraphics[width=0.9\linewidth]{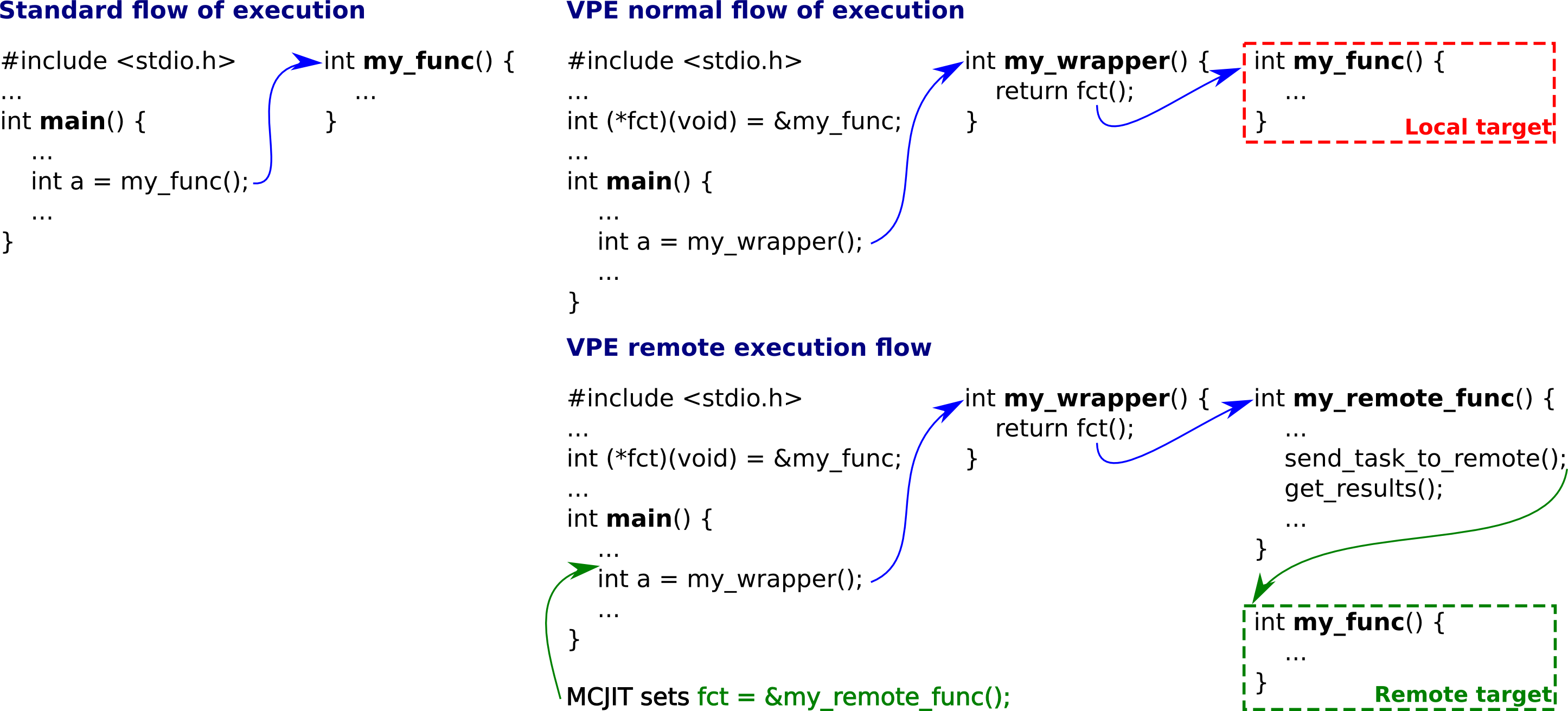}
\caption{Comparison of the execution flows in a standard system (left column) and in VPE (right column). While without VPE the JIT directly invokes the desired function,
  in VPE an intermediate step through a wrapper has to be made.
  When a remote target is selected, the wrapper invokes a function that is in charge of handling the communication with it, sending it the parameters and the code, and waiting for
  the results to be handed back.}
\label{fig:code_rempl}
\end{figure}

Invoking a function on the remote target is particularly tricky: while LLVM's MCJIT compiler includes a Remote Target Interface, it has the peculiarity of operating on modules only,
where a module is a collection of functions~\cite{Lopes14}.
This behaviour has only very recently been changed with the introduction of a new JIT, called ORC, but this code is still under development and it is available for the x86\_64 architecture only.
Operating at module level is very uncomfortable, as MCJIT requires a module to be finalized before being executed, and leaves us no simple way to alter the function invocation at run-time.
To acquire the capacity of dynamically dispatching functions, we thus automatically replace all functions with a caller that, in normal situations, simply executes the corresponding
function via a function pointer (see Fig.~\ref{fig:code_rempl}).
While this introduces a call overhead (as all function invocations must perform this additional ``caller step''),  when we wish to execute a function on the remote target, we just
have to alter this function pointer to make it point to another function that deals with the remote target, as shown in Fig.~\ref{fig:code_rempl}.

Similarly, when we consider that a function is not worth remote execution anymore --- for instance we might have observed that the remote target is slower that the local CPU on the given task, we
know that the remote target is already busy, or we have a more suitable function for the given computation unit --- we set back this pointer to its original value.
Since computing-intensive functions are automatically detected and offloaded to the remote target, the overhead imposed by the additional step quickly becomes negligible.
\vspace{-1em}

\subsection{Memory allocation problem}
\vspace{-.6em}
As briefly mentioned in the previous paragraph, once a function is invoked on the remote target, all data relative to it have to be
transferred as well.
In standard SMP systems this issue is not very relevant: all processors usually access all memory space, and
hardware mechanisms guarantee cache coherency.
On heterogeneous systems, however, the problem is more relevant.
Often the remote target has only a partial access to the main system's memory and there is no hardware support to ease the data sharing.
In this context differences between systems are remarkable, and we can distinguish between two macro-categories based on memory organization:
we have indeed systems with shared memory (physically shared or virtually shared), and systems without it.
We stress here that the two types can easily co-exist in the same platform, and while a sub-set of processing units can see the memory as a single address space, a different sub-set can
provide a different view.

In the context of VPE we consider only shared memory systems; in systems where this assumption is proven false, we could adopt a message passing layer to virtualize the real hardware
resources as in~\cite{Damschen15b}.
\vspace{-1em}

\section{Experimental Setup}
\label{sec:reptar}
\vspace{-1em}
To validate our proposal, we looked for an heterogeneous platform suitable for building a demonstrator.
We have chosen a TI-DM3730 DaVinci digital media processor SoC. It is present on the REPTAR platform~\cite{Dassatti14} we could use for our tests
and has already been adopted by~\cite{Nasrallah13}, allowing us an indirect comparison.
The DM3730 chip hosts an ARM Cortex-A8 1GHz processor and a C64x+ DSP processor running at 800MHz.
Part of the address space is shared between the two processors, therefore we can easily transfer data by placing them in this region.
This can be achieved by custom memory management functions, which however do not require any human intervention: when the JIT loads the IR code, it detects the memory operations and
automatically replaces them with our custom ones.
Note that this setup is not restrictive, as transfers among non-shared memory regions can be easily achieved by a framework such as MPI, as in~\cite{Damschen15b}.

The chosen DSP lacks an LLVM back-end we could use to automatically compile the code we are running in the JIT.
The TI compiler used to produce the binaries executed on the DSP is proprietary software, and writing a compatible back-end was out of the scope of the project.
While this could appear as a major obstacle, we have circumvented it by creating a set of scripts that compiles the functions' code using the aforementioned closed-source compiler,
and then extracts a symbol table that is loaded and used in VPE.
\vspace{-1em}

\section{Benchmarking}
\label{sec:bench}
\vspace{-1em}
\subsection{Methodology}
\vspace{-.6em}
We have evaluated the performance of VPE using a set of six algorithms: construction of the complementary nucleotidic sequence of an input DNA sequence, 2D
convolution with a square kernel matrix, dot product of two vectors, multiplication of two square matrices, search of a nucleotidic pattern in an input DNA sequence, and Fast Fourier Transform (FFT).
These algorithms were inspired by the Computer Language Benchmarks Game\footnote{\url{http://benchmarksgame.alioth.debian.org}} and were adapted to limit the use of floating point numbers, which are only
handled in software by the DSP we use and would, therefore, strongly penalize it.
The applications have been written in their naive implementation, that is, without any thorough handcrafted optimization\footnote{We will release the source code of the adopted tests after paper acceptance.}, and have been compiled on the ARM target with all the optimizations turned on (\emph{-O3}).
For each algorithm, a simple application allocates the data and calls the computing-intensive function repeatedly, in a continuous loop.
The size of the data is constant and the processing is made on the same data from one call to another.
The execution time of the processing function --- including the target selection mechanism, the call to the function, and the execution of the function itself --- is recorded at each iteration.

We have compared the performance of the algorithm running on the ARM core with the performance of the same algorithm on the DSP, once VPE has taken the decision to dynamically dispatch the
function to the DSP.
The performances reported for VPE skip this initial warm-up phase where the algorithm is first run on the CPU while VPE records the performances, as this value quickly becomes negligible as
the number of iterations of the algorithm increases.
\vspace{-1em}

\subsection{Results and analysis}
\begin{figure}[t]
  \centering
  \begin{tabular}{@{}ccc@{}}
    \hspace{-2mm} \vspace{0pt} \includegraphics[width=0.49\linewidth]{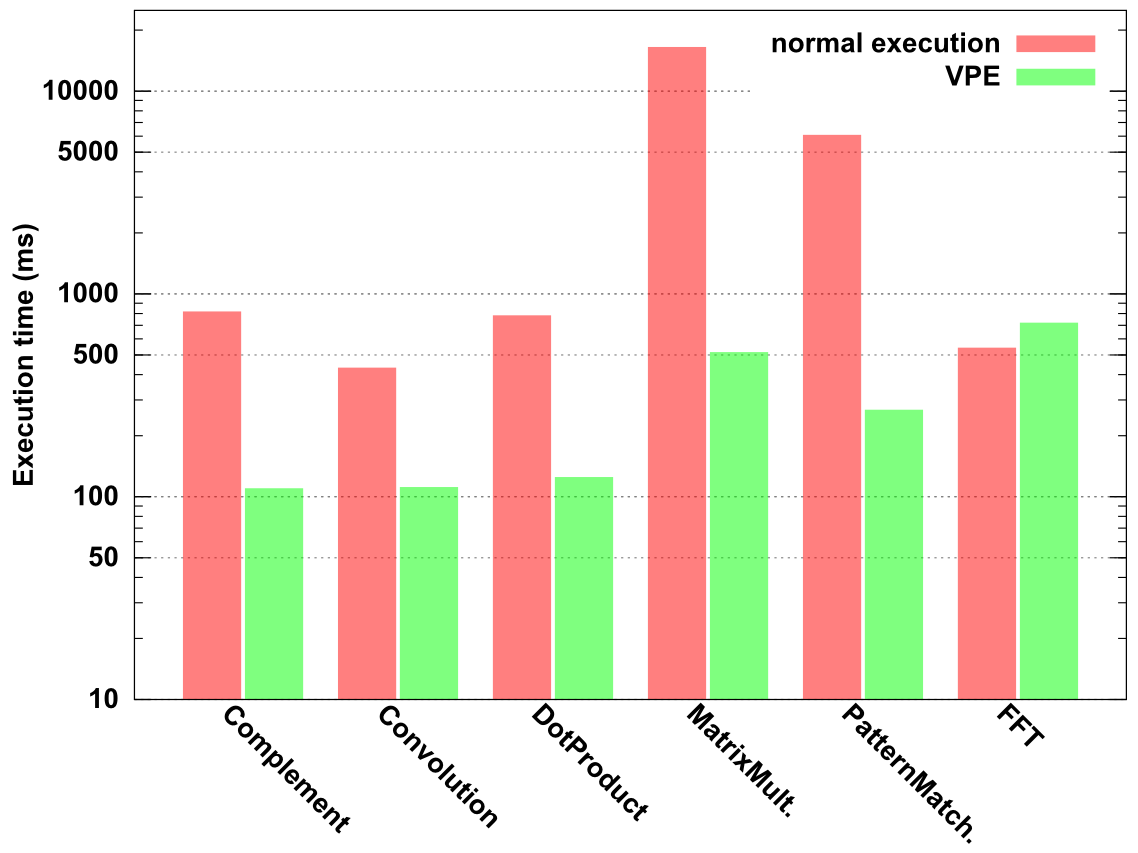} &
    \hspace{1mm} &
    \vspace{0pt} \includegraphics[width=0.49\linewidth]{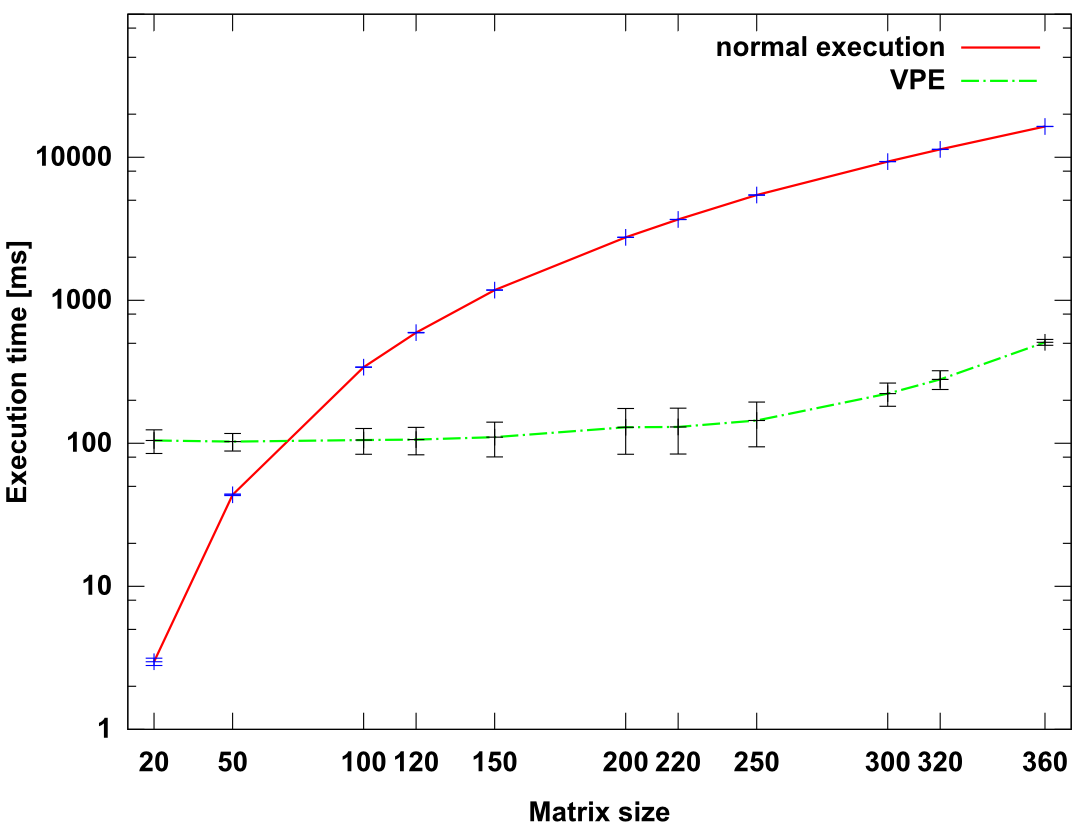} \\
    \bf{(a)} & & \bf{(b)}\\
  \end{tabular}
  \caption{{\bf(a)} Execution time of the algorithms on the REPTAR platform. The performance of the algorithm running on the ARM core and the performance of the same algorithm dispatched on the DSP,
    after the transition triggered by VPE. The execution times are given in milliseconds (the axis scale is logarithmic). {\bf(b)} Execution time of the matrix multiplication algorithm for a varying
    matrix size. Despite the ARM code being compiled with all the optimizations turned on (\emph{-O3}), the DSP largely outperforms it for matrices with size greater than $75 \times 75$.}
\label{fig:e_time}
\end{figure}

\vspace{-.6em}
Figure~\ref{fig:e_time}(a) shows that the execution time of the selected algorithms on the ARM core can be in the order of seconds.
This is notably the case for the matrix multiplication, but the other tests do not score far better.
Once VPE has selected the DSP as a remote target, noticeable improvements in terms of performance can be observed: the acceleration of the nucleotidic complement nearly reaches a factor
of eight, while the convolution sports a $4\times$ speedup.
Detailed timings for the different algorithms are reported in Tab.~\ref{tab:ex_time}.
Please note that the standard deviation is significantly increased when the code is running on the DSP under the control of VPE, since the profiler periodically slows down the execution while
collecting and analyzing usage statistics.

The most significant improvements have been obtained with the matrix multiplication and the pattern matching. Indeed, since the original versions of the algorithms are based on nested loops,
the TI compiler has detected optimization opportunities and carried out software pipelining that resulted in a reduction of the number of required CPU cycles, thereby increasing the execution
speed on the DSP target.
Figure~\ref{fig:e_time}(b) shows, on a logarithmic scale, the time required for matrix multiplication as a function of matrix size: for small matrices ($ < \sim 75 \times 75$), we can see that it
is not worth executing the operations on the DSP, as the time required for the setup (around 100ms) exceeds the execution time for the ARM processor --- although a remote execution
would still have, in this case, the advantage of freeing the CPU for other tasks. For bigger matrices, however, the advantage becomes considerable.
The versatility of our approach comes again handy in this case: we could easily, for instance, learn automatically a correlation between the size of the matrix passed as a parameter and the
performance achieved --- this could achieve this using a simple decision tree~\cite{Safavian91} ---, and ground future decisions upon this criteria.

While the improvements are remarkable, the optimization strategy we have selected --- that is, blindly off-loading the code to the DSP --- does not guarantee we have indeed a performance
improvement.
This is the case for the FFT code, which suffers a $25\%$ performance penalty from being executed on the DSP.
This performance penalty is due to the non-optimality of the code for the particular architecture: the hand-optimized DSP version of the same algorithm requires on average 109ms,
while the code executed by VPE takes around 720ms.
Two important points can be observed here: VPE will never be capable of outsmarting a developer in its job of optimizing the code for a particular architecture, and the optimization it performs
might not always be the best choice available.
For the former point, the result is a consequence of the fact that the code has been written without any knowledge of the system it will be executed upon, and thus it cannot benefit from the
system's peculiarities.
The improvements given by VPE, however, come ``for free'' from the application developer's stance since this result requires no effort from his side; This contrasts with, for instance,
the achievements of~\cite{Nasrallah13}.
Concerning the latter point, it is linked with the amount of knowledge available to VPE and to the amount of intelligence we have incorporated in it.
A more clever optimization strategy, as well as a better investigation of the type of operations performed inside the routine candidate for off-loading and a thorough analysis of the
statistics collected by perf, could have lead to a better choice --- which would have been, in the FFT case, leaving the FFT function on the ARM processor.
However, the dynamic nature of VPE forgives us these optimization attempts, as we can easily detect a mediocre performance on the remote unit and reverse our decision. This is an opportunity which is not available in, for instance, the work of~\cite{Damschen15a,Damschen15b}.
\vspace{-1em}

\begin{table}[t]
  \centering
  \caption{Timings (in ms) for the different algorithms tested on the REPTAR platform.
    The number reported after the $\pm$ represents one standard deviation.
    With ``normal execution'' we indicate the execution of the algorithm on the ARM CPU with no performance collection undergoing, while with ``VPE'' we indicate the very
    same code but running on the DSP in the VPE framework}
  \begin{tabular}{@{} lccr @{}}
    \toprule
    {\bf Algorithm} & {\bf normal execution} & {\bf VPE}       & {\bf Speedup} \\
    \midrule
    Complement      & $818.4 \pm 6$          & $109.9 \pm 29 $ & $\textcolor{blue}{  7.4\times}$ \\
    Convolution     & $432.2 \pm 1$          & $111.5 \pm 31 $ & $\textcolor{blue}{  3.8\times}$ \\
    DotProduct      & $783.8 \pm 1$          & $124.9 \pm 43 $ & $\textcolor{blue}{  6.3\times}$ \\
    MatrixMult.     & $16482.0 \pm 158$      & $515.9 \pm 35 $ & $\textcolor{blue}{ 31.9\times}$ \\
    FFT             & $542.7 \pm 1$          & $720.9 \pm 38 $ & $\textcolor{red} {  0.7\times}$ \\
    PatternMatch.   & $6081.7 \pm 58$        & $268.2 \pm 48 $ & $\textcolor{blue}{ 22.7\times}$ \\
    \hline
  \end{tabular}
  \label{tab:ex_time}
\end{table}

\subsection{Image processing prototype}
\vspace{-.6em}
We have also built a prototype demonstrator for the REPTAR board that uses a 2D convolution algorithm to detect contours in a video, similar in spirit to the one proposed in
SOSoC\footnote{\url{http://reds.heig-vd.ch/rad/projets/sosoc}}.
Both the CPU usage and the frame rate are displayed during the execution of the video processing.
We use the OpenCV library to decode and display the video frames in a dedicated process.
The system starts by invoking the video process that is in charge of decoding the current frame, then the pixel matrix is sent to the convolution process.
The computation of the convolution is performed within VPE and the resulting matrix is sent back to the video application, which displays it.

\begin{figure}[t]
  \centering
  \begin{tabular}{@{}cc@{}}
    \includegraphics[height=35mm]{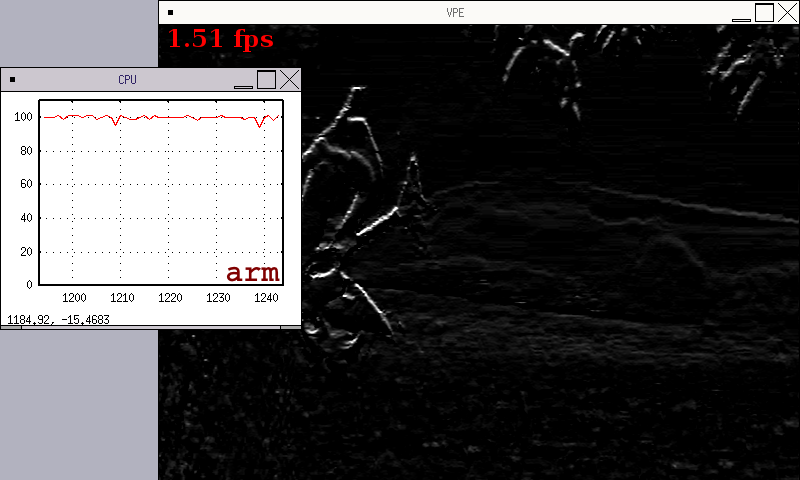} &
    \multirow{3}{*}[5.3em]{\includegraphics[width=0.5\linewidth]{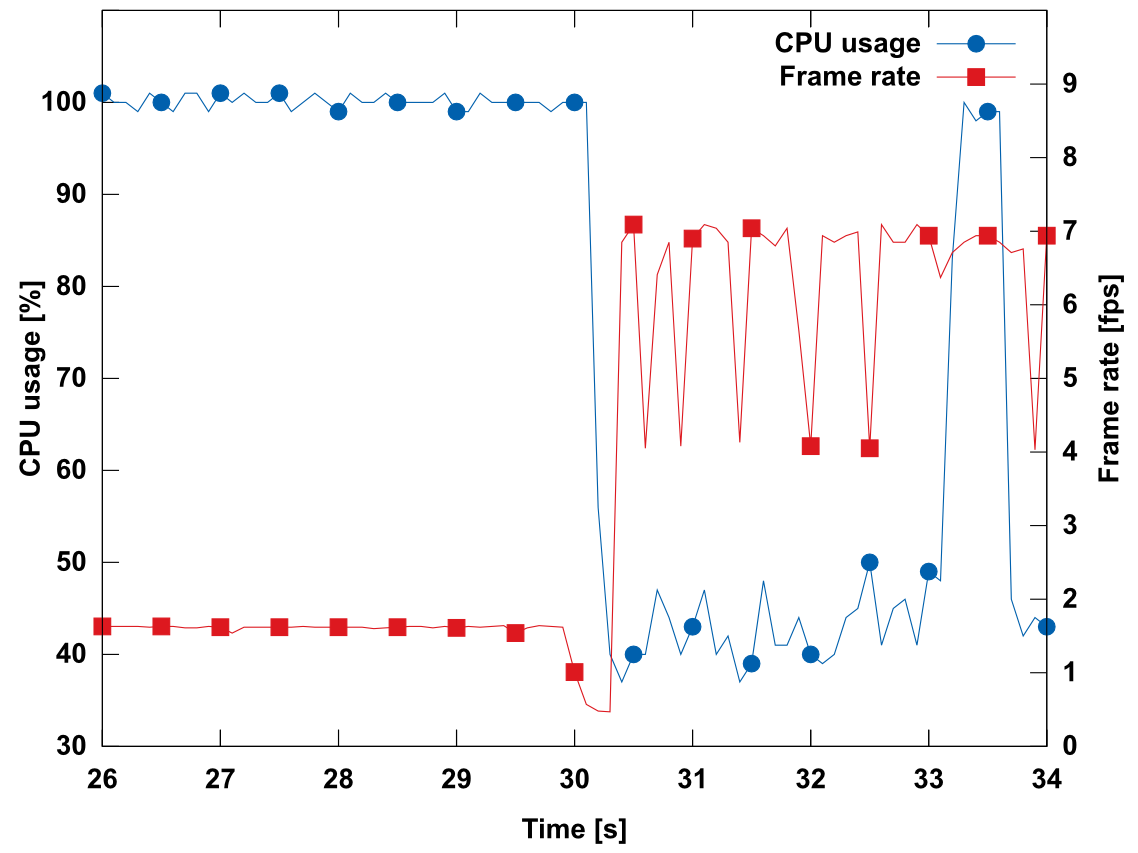}} \\
    \bf{(a)} & \\
    \includegraphics[height=35mm]{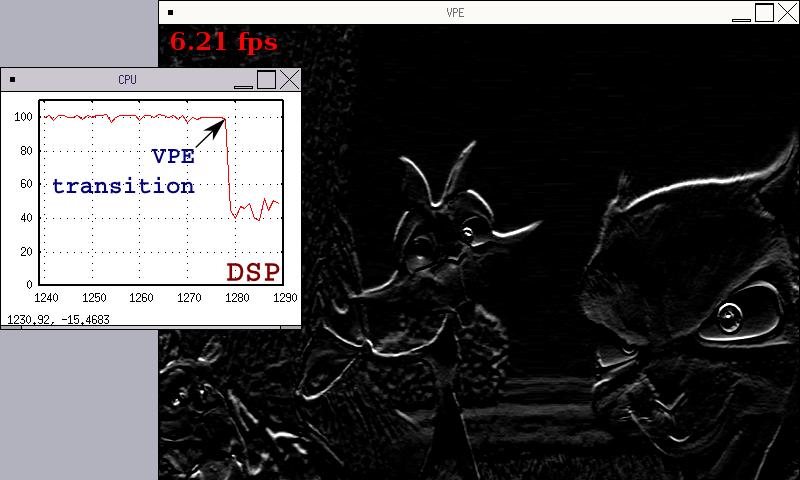} & \\
    \bf{(b)} & \bf{(c)}\\
  \end{tabular}
\caption{Screenshots of the VPE system in execution on the REPTAR platform. The system is working on a signal processing task --- contour detection in this case ---
  on a video while recording the percentage of CPU usage (in the small graph) and the frame rate (in the top-left corner of the video player).
  {\bf(a)} depicts the system before VPE transitioned the computation-intensive task (in this case, 2D convolution) to the DSP, while {\bf(b)} shows the system after this
  transition happened. It can be seen that when VPE triggers the transition on the DSP processor, the load of the ARM core is considerably relieved
  --- but still not negligible as the ARM core has to perform all the visualization-related tasks --- and the frame rate increases by a factor four.
  {\bf(c)} CPU usage and frame rate for the image processing prototype. The
  system starts on the CPU and performs statistics calculation. Then, when we allow it to change its execution target with a specific command, it decides to move the heaviest computation it is
  performing --- in this case a 2D convolution --- to the DSP, relieving the CPU load. At the same time, the frame rate gets multiplied by a factor four. Slightly after this moment,
  a rapid peak in CPU usage that is due to the performance calculation is visible.}
\label{fig:scr_vpe}
\end{figure}

Figure~\ref{fig:scr_vpe}(a) shows that, despite the main CPU being under heavy load, the frame rate is very low, being around 1.5fps.
After a predefined time interval, chosen to allow the spectators to observe the system running for a while, VPE is granted the right to automatically optimize the execution.
Once this happens, it detects that the convolution is the most expensive task and starts sending the new frames to the DSP, halving the CPU load --- the image handling is still performed
by the CPU --- and multiplying by four the frame rate.
Short bursts of CPU usage are, however, to be foreseen even when the convolution code is running on the DSP, as VPE still periodically analyzes the collected performances to spot variations in the
system's usage that could trigger a different resources allocation policy.
A detailed view of the CPU usage and frame rate evolution are shown in Fig.~\ref{fig:scr_vpe}(c).
\vspace{-1em}

\section{Conclusion}
\vspace{-1em}
\label{sec:Conclusion}
In this paper we have presented a transparent system optimization scheme capable of using a
run-time code profiler and a JIT to automatically dispatch computing-intensive chunks of code
to a set of heterogeneous computing units.
We have also built a working prototype that exploits this technique to accelerate by a factor of four
a standard image processing algorithm, and significantly improve the performances on a set of standard
benchmarks.

Future work will concentrate on testing our approach on a larger number of platforms, as well as
exploring additional run-time optimization schemes that could further reduce the algorithm's
computation time.
\vspace{-1em}

\bibliographystyle{splncs03}
\bibliography{paper}

\end{document}